\documentclass[twocolumn,showpacs,pre]{revtex4}%
\usepackage{amssymb}
\usepackage{amsmath}
\usepackage{graphicx}
\usepackage{dcolumn}
\usepackage{amsfonts}%
\providecommand{\U}[1]{\protect \rule{.1in}{.1in}}
\begin{document}
\title{Two-dimensional Classical Wigner Crystal: Elliptical Confining Potential}
\author{Zhen-zhong Zhang }
\author{Kai Chang}
\altaffiliation{Electronic address: kchang@red.semi.ac.cn}

\affiliation{NLSM, Institute of Semiconductors, Chinese Academy of Sciences, P. O. Box 912,
Beijing 100083, China}


\begin{abstract}
We investigate theoretically the ground-state configurations of
two-dimensional charged-particle systems with an elliptical hard-wall boundary
and their vibrational eigenmodes. The systems exhibit a series of structural
transitions, finally changing from a zigzag structure to a one-dimensional
Coulomb chain, as the eccentricity of the elliptical hard-wall boundary is increased.

\end{abstract}
\maketitle

\section{Introduction}

In 1934, Wigner predicted that electrons will crystallize in a
three-dimensional electron gas and form a lattice when the density of the gas
is lowered to a certain critical value. Since then, Wigner crystals have been
observed in various low-dimensional systems. These systems include electrons
on the surface of liquid helium \cite{PRL42-795}, electrons confined in
semiconductor quantum dots \cite{PRL68-3088}, colloidal suspensions with
macroscopic particles \cite{PRL82-3364}, vortices in superfluids
\cite{PRL68-3331}, and strongly coupled dusty plasma \cite{PRL72-4009}.

Recently, classical Wigner crystals have aroused considerable interest both
experimentally
\cite{PRE67-016411,PRE65-061405,PRL87-115002,EurophysLett55-45,PRE73-056404}
and theoretically
\cite{PRE68-060401,PRL88-125002,PRE69-036412,JPC10-11627,PRL70-818}.
Theoretical results for the ground-state configurations and normal mode
spectra of classical Wigner crystals agree with experiments very well. Very
recently, Melzer observed experimentally \cite{PRE73-056404} that a
two-dimensional (2D) charged-particle system under an anisotropic parabolic
confining potential forms a one-dimensional (1D) chain through a zigzag
transition, which agrees with the theoretical expectation
\cite{PRL70-818,JPC10-11627}. Most previous studies have focused on systems
under a parabolic confining potential or Coulomb potential induced by a fixed
positive charge. There have been only a few investigations of isotropic
hard-wall confinement \cite{PRE69-036412,PRB49-2667}. These have adopted Monte
Carlo simulation augmented with the Newton optimization technique. Since
Newton optimization cannot directly minimize the energy of a system under the
hard-wall confining potential, the authors \cite{PRE69-036412} adopted an
$r^{n}$-like confinement to simulate hard-wall confinement in order to study
the static and dynamical properties, because an $r^{n}$-like potential
approaches the hard-wall confining potential when the exponent $n$ approaches
the infinity.

In this paper, we adopt a nonlinear optimization algorithm (L-BFGS-B)
\cite{LBFGS} to investigate the ground-state configuration of a 2D
charged-particle system under an ellipitical hard-wall confining potential.
The excitation modes for different shapes of confining potential are also
studied. The present paper is organized as follows. In Sec. II, our model and
numerical method are introduced. In Sec. III, we calculate the ground-state
configurations and find that they exhibit a series of structural phase
transitions as the anisotropy of the elliptical hard-wall boundary is
increased. Sec. IV describes the influence of the anisotropy of the hard-wall
confining potential on the normal mode excitation spectrum. Finally, our
conclusions are given in Sec. V.

\section{Theoretical model}

The potential energy of a 2D system of $N$ charged particles including the
Coulomb potential under a hard-wall confining potential $V_{c}(r)$ is given
by
\begin{equation}
\Phi=\sum \limits_{i=1}^{N}(V_{c}(r_{i})+\frac{q^{2}}{\varepsilon}%
\sum \limits_{i>j}^{N}\frac{1}{\left \vert \overrightarrow{r}_{i}%
-\overrightarrow{r}_{j}\right \vert }) \label{1}%
\end{equation}

In the case of an elliptical hard-wall confining potential, utilizing the
coordinate transform $x_{i}=r_{i}\cos \theta_{i},y_{i}=er_{i}\sin \theta_{i}$,
where $e$ refers to the eccentricity of the ellipitical boundary, the
elliptical hard-wall confining potential can be expressed as
\begin{equation}
V_{c}(r)=\{%
\begin{array}
[c]{c}%
0\text{ if }r<R_{c}\\
\infty \text{ if }r\geq R_{c}%
\end{array}
, \label{2}%
\end{equation}
where $R_{c}$ is the major axis length.

By choosing $r_{0}=R_{c}$ as the unit of length and $E_{0}=q^{2}/(\varepsilon
R_{c})$ as the unit of energy, the potential energy in dimensionless form is
\begin{equation}
\Phi \!=\! \sum \limits_{j>i}^{N}\! \frac{1}{\sqrt{(r_{i}\cos \theta
_{i}\!-\!r_{j}\cos \theta_{j})^{2}\!+\!e^{2}(r_{i}\sin \theta_{i}\!-\!r_{j}%
\sin \theta_{j})^{2}}} \label{3}%
\end{equation}

where $0\leq r_{i},r_{j}<1,0\leq \theta_{i,}\theta_{j}\leq2\pi.$ From this
form, we note that the Hamiltonian depends only on the number of particles $N$
and the eccentricity $e$ of the ellipitical boundary. Due to the invariance of
the system configuration after a $\pi/2$ rotation, we only investigate the
cases of $0\leq e\leq1$ in this paper.

The L-BFGS-B algorithm \cite{LBFGS}, which is \ an improved version of the
BFGS \cite{Numerical Recipes} algorithm, can find the minimum of a function
$f[r(x_{1},x_{2},\cdots x_{i}\cdots)]$ under the constraint of $l_{i}\leq
x_{i}\leq u_{i}.$ To solve the boundary problem, we obtain the piece-wise
linear path $x(t)=P\left(  x_{i}-tb_{i},l,u\right)  $ by projecting the
steepest descent direction onto the feasible region:%

\begin{equation}
P(x,l,u)=\left \{
\begin{array}
[c]{ccl}%
l_{i} & \text{if} & x_{i}<l_{i}\\
x_{i} & \text{if} & x_{i}\in \lbrack l_{i},u_{i}]\\
u_{i} & \text{if} & x_{i}>u_{i}%
\end{array}
\right.  . \label{10}%
\end{equation}

Such a transformation enables us to find the local minimum of the univariate
piece-wise quadratic $q_{n}(t)=y_{n}(x(t))$, in which $x^{c}$ is defined as
the Cauchy point. The variables whose value at $x^{c}$ is at the lower or
upper bound, comprising the active set $A(x^{c})$, are held fixed. Then we
consider the following problem:%

\begin{align}
min\{f(x  &  :x_{i}=x_{i}^{c},\  \  \forall i\in A(x^{c})\} \label{11}\\
l_{i}  &  \leq x_{i}\leq u_{i}\  \forall i\notin A(x^{c}). \label{12}%
\end{align}

First, adopting the conjugate gradient algorithm \cite{Numerical Recipes} on
the subspace of free variables, we approximately solve Eq. (\ref{11}). In the
iterative process, we take $x^{c}$ as the initial value for iteration, and
then truncate the path to satisfy Eq. (\ref{12}). After an approximate
solution $\overline{r}_{n+1}$ has been obtained, $f$ decreases further along
$d_{n}=\overline{r}_{n+1}-r_{n}$ on the condition that $r_{n+1}$ is in the
appropriate range (boundary constraint). Then the approximation Hessian matrix
and the gradient vector at $r_{n+1}$ are calculated, and a new iteration
starts. Such iterations continue until the required accuracy is achieved. More
detailed informaton about this algorithm can be found in Ref. \cite{LBFGS}.

The total energy of the system under the elliptical hard-wall confining
potential (see Eq. (\ref{3}) ) can be minimized by the L-BFGS-B algorithm.
First, all particle coordinates are generated randomly. Secondly, the L-BFGS-B
algorithm is adopted to minimize the total energy of the system. We repeat the
above procedure many times. Finally, we choose the configuration with the
lowest energy as the ground-state of the system. To check the accuracy and
validity of this algorithm, we reproduced the ground-state configurations for
an isotropic hard-wall potential \cite{PRE69-036412}.

Once the system ground-state is obtained, the vibrational eigenmode can be
calculated by diagonalizing the dynamical matrix (see Appendix). The
eigenfrequencies in this paper are expressed in the unit $\omega_{0}%
=\sqrt{E_{0}/(mr_{0}^{2})}$.

\section{Numerical Results and discussions}

\subsection{The ground-state configuration}

\begin{figure}[ptb]
\includegraphics[width=\columnwidth]{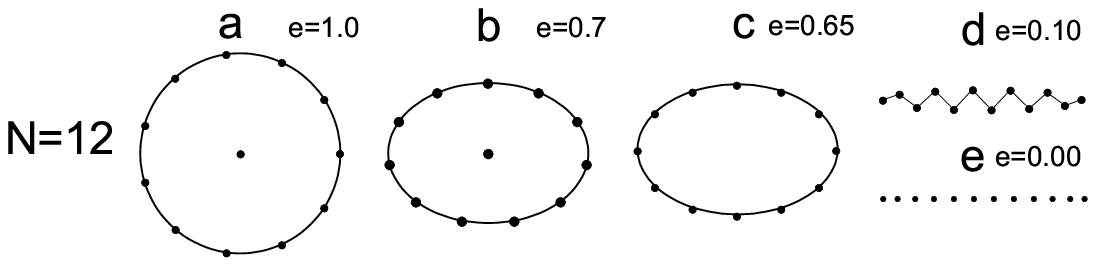}
\includegraphics[width=\columnwidth]{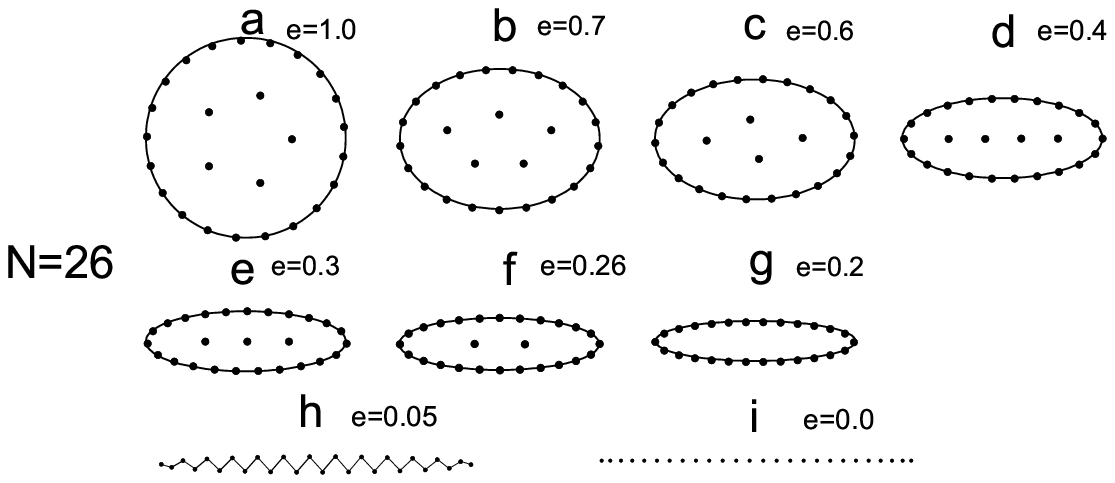}
\includegraphics[width=\columnwidth]{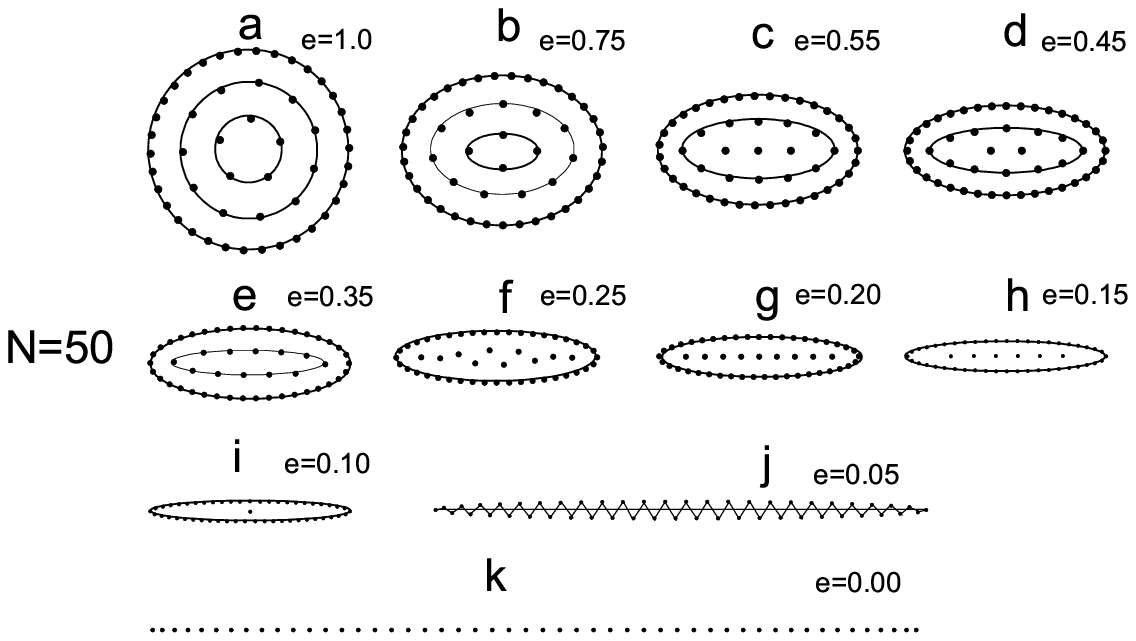}\caption{The ground state
configurations of three clusters with different numbers of particles $N=12$,
$26$, $50$ fors different eccentricities of the elliptical boundary.}%
\end{figure}

When the charged particles are confined by a hard-wall potential, they repel
each other due to the Coulomb interaction so that some particles first move to
the boundary until the outer shell is fully occupied, the remaining particles
stay in the inner shells. The shape of the boundary significantly affects the
system ground-state configuration. Fig.1 shows the evolution of the
ground-state configurations for different numbers of particles as the boundary
eccentricity $e$ decreases. The ground-state configurations in the anisotropic
case are quite different from those in the isotropic case ($e=1$). They change
significantly with decreasing eccentricity $e$, and exhibit zigzag structures
when $e$ approaches zero.

Taking the $N=50$ cluster as an example, we can find a continuous deformation
from circular to elliptical rings as the eccentricity $e$ decreases. The inner
ring collapses into a line with decreasing $e$. Notice that more and more
particles leave the central line and merge into the adjacent outer ring. Then
the system forms two elliptical rings (see Fig. 1). This process repeats until
the system exhibits a quasi-one-dimensional zigzag configuration and becomes a
straight line at $e=0$. In addition, inner shells are more sensitive to the
change of the eccentricity $e$ than outer shells. Decreasing $e$ further,
particles in the inner shell move to the outer shell, and the inner shell
finally collapses into a line. \begin{figure}[ptb]
\includegraphics[width=\columnwidth]{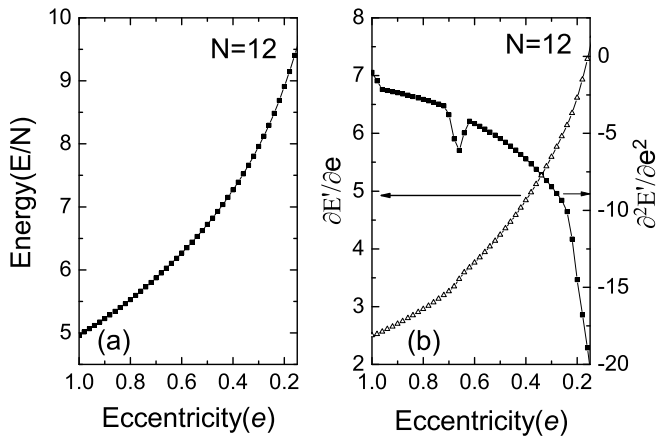}\caption{The ground-state
energys , the first derivative and second derivative of the ground state
energy with respect to the eccentricity as a function of the eccentricity for
$N=12$.}%
\end{figure}

We find that there are two different kinds of transitions: slow and abrupt
transitions. For convenience, we refer to the configuration in which $N-1$
particles are on the outer shell and one at the centre as $\alpha$-type, and
the configuration in which all particles are on the outer shell as $\beta
$-type. From the Table I in Ref. \cite{PRE69-036412}, we know that the system
exhibits the $\alpha$-type configuration for $N=12$ under a circular hard-wall
confining potential. But for an ellipitical confining potential, the
transition from $\alpha$-type to $\beta$-type occurs as the boundary
eccentricity $e$ decreases to a certain value. Fig. 2(a) shows the system
energy of $N=12$ as a function of the eccentricity $e$ when the major axis is
fixed and the minor axis decreases with $e$. When $e$ decreases and the area
is compressed, the total energy increases. We adopt $E^{\prime}=E\cdot
e^{1/2}$ and calculate the first and second derivatives of $E^{\prime}$ with
respect to $e$, i.e. $\partial E^{\prime}/\partial e,\partial^{2}E^{\prime
}/\partial^{2}e$, which are shown in Fig. 2(b). An obvious dip in
$\partial^{2}E^{\prime}/\partial^{2}e$ appears with the decrease of $e$. As
seen in Fig. 1, the system ground-state is $\alpha$-type at $e=0.7$ and
$\beta$-type at $e=0.65$. The transition from $\alpha$-type to $\beta$-type
takes place at $e\approx0.666$, $0.546$, and $0.483$ respectively for $N=12$,
$13$, and $14$. \begin{figure}[ptb]
\includegraphics[width=\columnwidth]{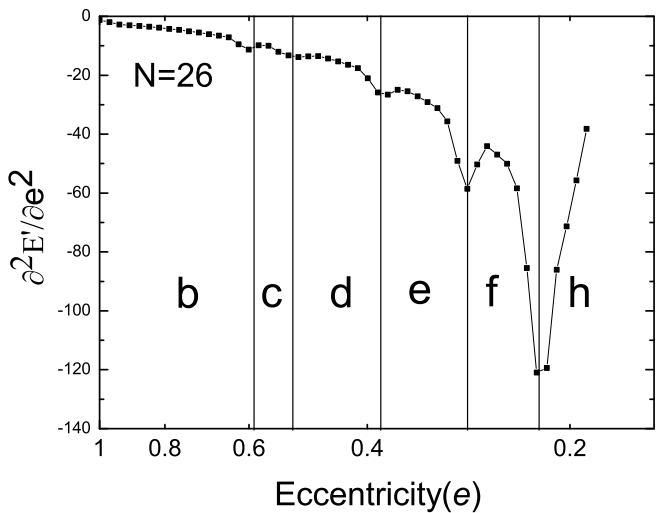}\caption{The second-order
derivative $\partial^{2}E^{\prime}/\partial e^{2}$ as a function of
eccentricity for $N=26$, where b-h correspond to the configurations for $N=26$
in Fig.1 }%
\end{figure}

In Figs. 3 and 4, we plot the second derivative of energy with respect to the
eccentricity $e$ for $N=26$ and 50 as a function of eccentricity $e$. More
transitions, which are denoted by the vertical lines, can be found.
\begin{figure}[ptb]
\includegraphics[width=\columnwidth]{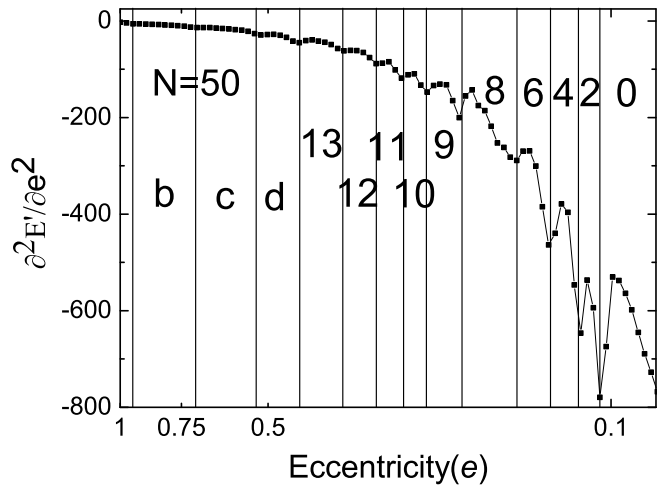}\caption{Same as Fig.3, but
for $N=50$. The numbers denote the number of particles in the inner shell. }%
\end{figure}From the evolution of three clusters, the ground-state
configuration finally transits from $\alpha$-type to $\beta$-type and finally
to a 1D chain through a zigzag structure. Note how the system evolves from 2D
to 1D. In Fig. 5(a) and 5(b), we plot zigzag angles and particle density of
zigzag structures and a 1D chain as a function of particle position for
$N=50$. From Fig. 5(a), we find the density per unit length of a 1D chain
($e=0$) exhibits behavior similar to that of the zigzag structure as the $x$
coordinate increases. It remains almost constant in the central region and
increases significantly at the edges of the chain. These behaviors of a 1D
chain under hard-wall confinement are very different from those under a
parabolic potential, which was investigated by Dubin. \cite{PRE55-4017} and
Melzer \cite{PRE73-056404}. \begin{figure}[ptb]
\includegraphics[width=\columnwidth]{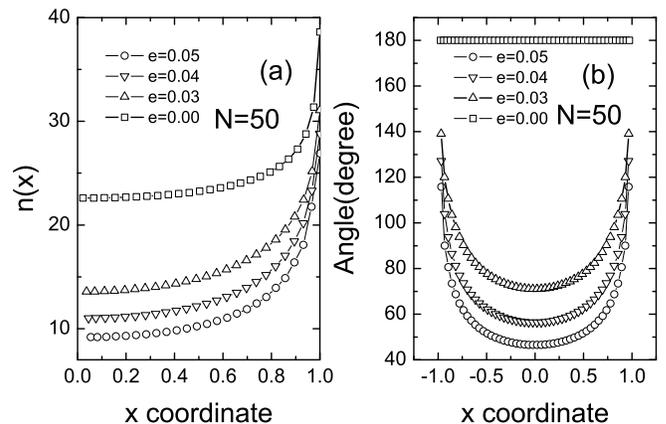}\caption{Density per unit
length $n(x)$, and zigzag angle as a funtion of coordinate $x$ for different
eccentricities.}%
\end{figure}

\section{The eigenmode spectrum}

\begin{figure}[ptb]
\includegraphics[width=\columnwidth]{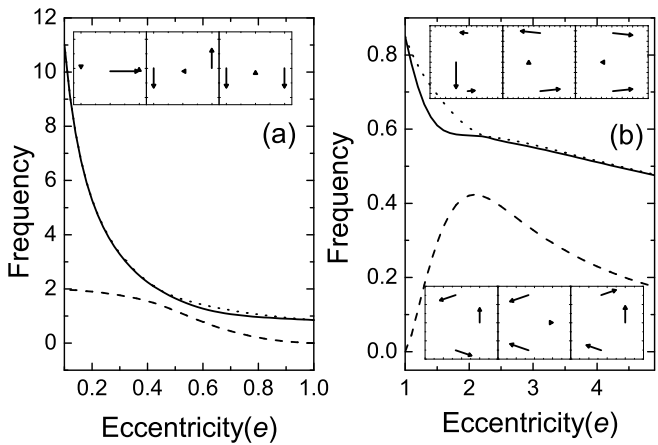}\caption{The frequency
spectra as functions of the eccentricity $e$ for a cluster $N=3$. The dashed,
solid, and dotted lines denote the frequencies of three vibrational modes. The
eigenvectors, with frequencies arranged in ascending order from left to right,
are in the insets in Fig. 6(a) for $e=0.1$. The eigenvectors of $e=1.6$ are in
the lower insets of Fig. 6(b), and those of $e=2.8$ are in the upper insets of
Fig. 6(b).}%
\end{figure}

In this section, we discuss the vibrational modes of the ground state under
the elliptical hard-wall confining potential.

Because the radial motion of the particles at the boundary is practically
frozen, only $2N-N_{E}$ modes exist when the $N_{E}$ particles are located at
the boundary. But under ellipse hard-wall confinement, the rotational mode
with zero frequency at $e=0$ becomes nonzero and some degenerate modes split
due to the breaking of cylindrical symmetry. In addition, increasing the
eccentricity $e$ results in abrupt transitions of the ground-state
configurations ($N\geq12$), and their eigenmodes change significantly because
of the abrupt change of the number of particles at the boundary.

\subsection{$N=3$ particle cluster}

\begin{figure}[ptb]
\includegraphics[width=\columnwidth]{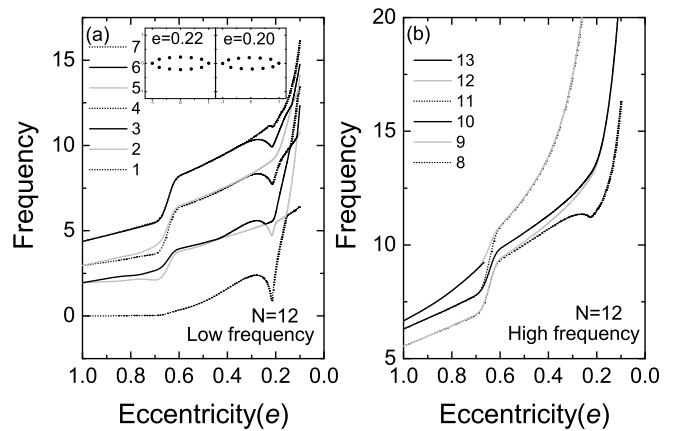}\caption{The frequency
spectra as funtions of the eccentricity $e$ for a cluster $N=12$.}%
\end{figure}

The frequencies of three vibrational modes in an $N=3$ cluster as a function
of eccentricity $e$ are shown in Figs. 6(a) and 6(b). One can find two
different mechanisms driving the transition of the ground state configuration
as the anisotropy increases. One is due to anisotropic hard-wall confinement.
The other is the minor (major) axis of the ellipse decreases (increases), and
consequently the Coulomb interaction increases (decreases) for $e<1$ ($e>1$).
When $e$ approaches 1, the former mechanism is dominant and the rotational
mode frequency increases (see dashed line in Fig. 6(b)). In addition, this
mechanism make high-frequency (HF) degenerate mode split into two modes (see
Fig.6(a,b) solid and dotted lines). When $e$ decreases (increases) further,
the latter mechanism becomes more important, and two split HF modes degenerate
approximately again and the low-frequency (LF) mode saturates gradually. From
the eigenvectors at $e=0.1$ (see the insets of Fig. 6(a)), we find that
decreased eccentricity $e$ enhances the confinement along the $y$ axis, and
consequently leads to the increases of frequencies of the vibration modes
along the $y$ axis. Thus those frequencies tend to infinity in the extreme
case of $e=0$. From the 1D Coulomb chain potential function, we diagonalize
the dynamical matrix of the ground-state configuration for an $N=3$ cluster
and get that the lowest frequency is 2.0 and the two high frequencies are
infinity, which agrees with Fig. 6(a). In addition, Figs. 6(a) and 6(b) show
that the two split modes cross respectively at $e\approx2.3$ and $0.44$
(notice $2.3\approx1/0.44$). We compare the eigenvector of the highest
frequency at $e=1.6$ with that at $e=2.8$ (see the insets of Fig.6(b)). One
can see that two particles near the major axis move in the same direction at
$e=1.6$ but in the opposite direction at $e=2.8$. This verifies that the cross
occurs indeed between $e=1.6$ and $e=2.8$.

\subsection{$N=12$ particle cluster}

In this subsection we discuss a larger system ($N=12$) to illuminate the
effect of the elliptical hard-wall boundary on the spectrum of the system.
Fig. 7 shows all eigenfrequencies of the $N=12$ cluster as a function of the
eccentricity $e$. From this figure, we see that the frequency spectra exhibit
many jumps between $e=0.668$ to $e=0.6$. This is because the system transits
from $\alpha$-type to $\beta$-type at $e\approx0.666$ (see the definition of
$\alpha$-type and $\beta$-type in Section II). The distance between
neighboring particles decreases, resulting in an abrupt increase of the
eigenfrequency. In addition, the highest frequency mode diminishes due to the
freezing of the radial motion of all particles at the boundary (see the
thirteenth\ mode in Fig. 7(b)). It is interesting to note that a small dip
appears at $e\approx0.22$ which cannot be found in the energy figure of the
system (see Fig. 2). Checking the ground-state configuration around $e=0.22$,
we find that the system changes from $\beta$-type to the zigzag structure (see
the insets in Fig. 7(a)). Next, we explain the effect of the elliptical
boundary on vibrational modes in detail in three regions.

When $e\gtrsim0.66$, the system ground state configuration is $\alpha$-type,
and there are thirteen vibration modes. Among these modes, the anisotropic
effect results in the apparent splitting of the LF modes, but the effect on
the other higher frequency modes is negligible. \begin{figure}[ptb]
\includegraphics[width=\columnwidth]{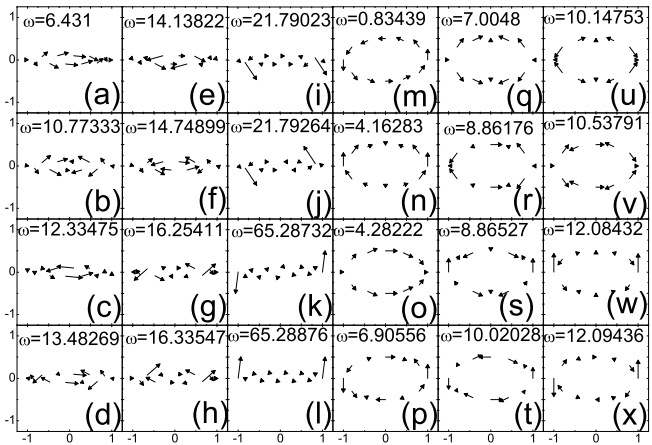}\caption{All eigenvector for
the clusters with $N=12$ particles at $e=0.1$ and $e=0.5$. (a-l) is for
$e=0.1$ and (m-x) is for $e=0.5$}%
\end{figure}

When $0.22\lesssim e\lesssim0.66$, the system ground state configuration is
$\beta$-type. Because the number of particles increases by one at the
boundary, and the HF modes recompose due to the configuration transition from
$\alpha$-type to $\beta$-type. As the eccentricity $e$ decreases, the second
mode crosses with the third mode (see Fig. 7(a)). A similar feature was
observed experimentally in an anisotropic parabolic confinement
\cite{PRE73-056404}. In the HF part (see Fig. 7(b)), the ninth mode departs
from the eighth mode and becomes degenerate with the tenth mode gradually,
corresponding to the slow transition of the configuration.

When $e\lesssim0.22$, the system ground-state configuration is a zigzag
structure. From the inset in Fig. 7, we find that the zigzag structure
destroys the configuration symmetry, resulting in the splitting of the
degenerate modes in the isotropic hard-wall confining potential. The lowest
vibrational mode is no longer the rotation mode, and more crosses between LF
modes (see Fig. 7(a)) can be found. The LF eigenvectors at $e=0.1$ and $e=0.5$
show clearly the crosses of different modes. From the eigenvectors of the
$N=12$ cluster in Fig. 8, we can compare the vibration direction of eigenmodes
at $e=0.1$ with that at $e=0.5$ and find that all corresponding particles move
in the same way among the vibration modes (a) and (o), (b)and (q), (c) and
(m), (d) and (r). The particles with larger vibration amplitudes in the LF
modes appear close to the center of the system as $e$ decreases. But the
behavior of the HF modes is opposite to that of the LF modes (see Fig. 8).

\section{Conclusion}

The ground-state configuration and dynamical properties of a 2D classical
cluster of charged particles in an elliptical hard-wall confining potential
are investigated theoretically. The dependence of the ground-state
configuration on the number of particles and the eccentricity of the confining
potential are discussed in detail. As the eccentricity $e$ decreases, the
system exhibits two different kinds of transitions, i.e., slow and abrupt
transitions. When increasing the system anisotropy, circular shells are
deformed into ellipses continuously, the inner shell collapses into a line,
and then number of particles in the line decreases gradually. Last, the system
exhibits a zigzag structure and forms a 1D Coulomb chain when $e$ approaches zero.

The boundary anisotropy significantly influence the frequency of the vibration
modes. With increasing the anisotropy of the boundary, the rotation mode
becomes nonzero and degenerate modes split due to the breaking of cylindrical
symmetry. The rotation mode is no longer the lowest vibration mode when the
eccentricity $e$ approaches zero.

\begin{acknowledgments}
This work was supported by the NSFC Grant No. 60376016, 60525405.
\end{acknowledgments}

\appendix*

In this appendix, we present the dynamical matrix of a 2D classical system
under the elliptical coordinates. Using the coordinate transformation,
$x_{i}=e\rho_{i}\cos \theta_{i},y_{i}=\rho_{i}\sin \theta_{i}$, where $e$
denotes the eccentricity of the elliptical hard-wall boundary, one can obtain
$\dot{x}=e\dot{\rho}\cos \theta-(\rho \sin \theta)\dot{\theta}$ and \ $\dot
{y}=\dot{\rho}\sin \theta+(\rho \cos \theta)\dot{\theta}$. The equilibrium
positions of the particles are denoted by $\left \{  \rho_{1}^{0},\cdots
\rho_{i}^{0},\theta_{1}^{0},\cdots \theta_{i}^{0}\right \}  .$

In the harmonic approximation around the equilibrium position, the Lagrangian
can be expressed as $L(\left \{  \rho_{i}\right \}  ,\left \{  \theta
_{i}\right \}  )=\frac{1}{2}m\sum \limits_{i}(\dot{\rho}_{i}^{2}(e^{2}\cos
^{2}\theta_{i}+\sin^{2}\theta_{i})+\rho_{i}^{2}(e^{2}\sin^{2}\theta_{i}%
+\cos^{2}\theta_{i})\dot{\theta}_{i}^{2}+2\rho_{i}(1-e^{2})\sin \theta_{i}%
\cos \theta_{i}\dot{\rho}\dot{\theta})-\frac{1}{2}\sum \limits_{ij}(\Phi
_{\rho_{i},\rho_{j}}\delta \rho_{i}\delta \rho_{j}+2\Phi_{\rho_{i},\theta_{j}%
}\delta \rho_{i}\delta \theta_{j}+\Phi_{\theta_{i},\theta_{j}}\delta \theta
_{i}\delta \theta_{j}),$ where $\delta \rho_{i}=\rho_{i}-\rho_{i}^{0},$
$\delta \theta_{i}=\theta_{i}-\theta_{i}^{0},$ $\Phi_{\rho_{i},\rho_{j}}%
=(\frac{\partial^{2}\Phi}{\partial \rho_{i}\partial \rho_{j}})_{eq},$
$\Phi_{\theta_{i}\theta_{j}}=(\frac{\partial^{2}\Phi}{\partial \theta
_{i}\partial \theta_{j}})_{eq},$ etc. Since $\delta \theta_{i},\delta \rho
_{i}\rightarrow0,$ linearizing the equation of motion for $\left \{  \rho
_{i}\right \}  $ and $\left \{  \theta_{i}\right \}  $ and defining
$T=(\delta \rho,\rho^{0}\delta \theta)$, we can obtain the equation of the
particles' movement under matrix form $\frac{d^{2}}{dt^{2}}T=-M\cdot T$. So
the dynamical matrix of a 2D classical system $M$ is $\left[
\begin{array}
[c]{cc}%
\frac{B\cdot G}{A\cdot C-B\cdot D}-\frac{C\cdot E}{A\cdot C-B\cdot D} &
\frac{B\cdot H}{A\cdot C-B\cdot D}-\frac{C\cdot F}{A\cdot C-B\cdot D}\\
\frac{A\cdot G}{B\cdot D-A\cdot C}-\frac{B\cdot E}{B\cdot D-A\cdot C} &
\frac{A\cdot H}{B\cdot D-A\cdot C}-\frac{B\cdot F}{B\cdot D-A\cdot C}%
\end{array}
\right]  $, where $\left[  A_{i,i}\right]  =(e^{2}\cos^{2}\theta_{i}^{0}%
+\sin^{2}\theta_{i}^{0}),$ $[B_{i,i}]=(1-e^{2})\sin \theta_{i}^{0}\cos
\theta_{i}^{0},$ $[C_{i,i}]=(e^{2}\sin^{2}\theta_{i}+\cos^{2}\theta_{i}),$
$[D_{i,i}]=(1-e^{2})\sin \theta_{i}^{0}\cos \theta_{i}^{0},$ $[E_{i,j}%
]=-\Phi_{\rho_{i},\rho_{j}},$ $[F_{i,j}]=-\frac{\Phi_{\rho_{i},\theta_{j}}%
}{\rho_{j}^{0}},$ $[G_{i,j}]=-\frac{\Phi_{\theta_{i},\rho_{j}}}{\rho_{i}^{0}}%
$, and $[H_{i,j}]=-\frac{\Phi_{\theta_{i},\theta_{j}}}{\rho_{i}^{0}\rho
_{j}^{0}}$.

\end{document}